\documentclass[twocolumn,amsmath,amssymb,widetext]{revtex4}
\usepackage{graphicx}

\begin{document}
\title{Surface acoustic wave-induced electroluminescence intensity oscillation in planar light-emitting devices}
\author{Marco Cecchini}
\email{cecchini@sns.it}
\author{Vincenzo Piazza}
\author{Fabio Beltram}
\affiliation{NEST-INFM and Scuola Normale Superiore, I-56126 Pisa,
Italy}
\author{D. G. Gevaux}
\author{M. B. Ward}
\author{A. J. Shields}
\affiliation{Toshiba Research Europe Limited, Cambridge Research 
Laboratory, 260 Cambridge Science Park, Milton Road, Cambridge CB4 OWE, United Kingdom}
\author{H. E. Beere}
\author{D. A. Ritchie}
\affiliation{Cavendish Laboratory, University of Cambridge,
Cambridge CB3 0HE, United Kingdom}


\begin{abstract}
Electroluminescence emission from surface
acoustic wave-driven light-emitting diodes (SAWLEDs) is
studied by means of time-resolved techniques. We show
that the intensity of the SAW-induced electroluminescence 
is modulated at the SAW frequency ($\sim 1$ GHz),
demonstrating electron injection into the p-type region
synchronous with the SAW wavefronts. 
\end{abstract}


\maketitle

Surface-acoustic-wave (SAW) based devices are nowadays 
widely used for mobile and wireless applications, as well 
as for satellite communications and military applications\cite{CampbellBook}. The large-scale 
diffusion of sophisticated communication systems stimulated a fast rise in the
SAW-based device market, reaching a production 
volume of several million devices per day\cite{RuppelBook}.
More recently SAWs have attracted the interest of the 
semiconductor community in view of the exploitation of their 
interaction properties with two-dimensional-electron-gases 
(2DEGs) embedded in semiconductor 
heterostructures\cite{wixforth-prl86,willet-prl93}. SAWs 
propagating through mesas containing high-quality 2DEGs indeed 
drive modifications on the 2DEG equilibrium state. 
Acoustic waves propagating along piezo-electric substrates 
are accompanied by potential waves which can trap electrons 
in their minima and induce dc currents or 
voltages\cite{wixforth-surfsci94,wixforth-ssc92,campbell-ssc92}. 
The discovery of the so-called acoustoelectric effect was 
followed by the proposal of innovative device concepts.
Among these Talyanskii \emph{et al.} proposed the 
implementation of a novel current standard, demonstrating 
very precise acoustoelectric current quantization due to 
charge drag by SAWs through a quantum point contact
\cite{cunningham-prb00,cunningham-prb99,ebbecke-apl00,
talyanskii-prb97,talyanskii-jpcm96}. 
Control over the constriction width allows very precise selection of 
the number of electrons packed in each SAW minimum
down to the single-electron-transport regime. One of the most 
appealing applications proposed after the first report of 
the acoustoelectric quantized current was to incorporate 
single-electron SAW pumps in planar 2D electron/2D hole 
gas (n-p) junctions to fabricate high-repetition-rate 
single-photon sources. 

Very recently two different groups demonstrated one of the 
main building blocks of such source, that is the possibility 
to switch on the electroluminescence of planar n-p junctions 
by applying SAWs\cite{hosey-apl04,cecchini-apl04}. From a 
more fundamental physics point of view these results
extend the acoustoelectric effects to planar systems where 
both electron and holes are present and allow detailed 
studies of the effects of the acoustic modulation in 
light-emitting devices (LEDs). 

In this Letter we report on time-resolved studies of
the acoustoelectric effects in 
planar LEDs. We fabricated devices containing lateral n-p 
junctions and interdigital transducers for SAW generation (SAWLEDs) 
and studied the optical properties of the 
devices induced by the acoustic perturbation. We monitored
SAW-induced electro-luminescence (EL) as a function of time 
and demonstrated the presence of oscillations with the same 
period of the SAW ($\sim 1$ ns).

Devices were fabricated starting with a p-type modulation-doped
Al$_{0.3}$Ga$_{0.7}$As/GaAs heterostructure grown by molecular
beam epitaxy, containing a two-dimensional hole gas (2DHG) within 
a 20-nm-wide GaAs quantum
well embedded 70\,nm below the surface. The measured hole density
and mobility after illumination at 1.5\,K were $2.0\times
10^{11}$\;cm$^{-2}$ and 35000\;cm$^{2}$/Vs, respectively. 

The heterostructure was processed into mesas with an annealed p-type
Au/Zn/Au (5/50/150\,nm) Ohmic contact. 
The fabrication of the 
n-type region of the junction followed Ref.\,\cite{cecchini-apl03}. The
Be-doped layer was removed from part of the mesa by means of wet etching
(48\,s in H$_{3}$PO$_{4}$:H$_{2}$O$_{2}$:H$_{2}$O = 3:1:50) and 
the sample was loaded into a thermal evaporator 
for the deposition of a self-aligned Ni/AuGe/Ni/Au (5/107/10/100\,nm)
n-type contact. After annealing, donors introduced by the n-type
contact provide conduction electrons within the well thus creating 
an electron gas below the metal pad, adjacent to the 2DHG (inset of Fig\,1(a)).
\begin{figure}[!ht]
\includegraphics[width=8cm]{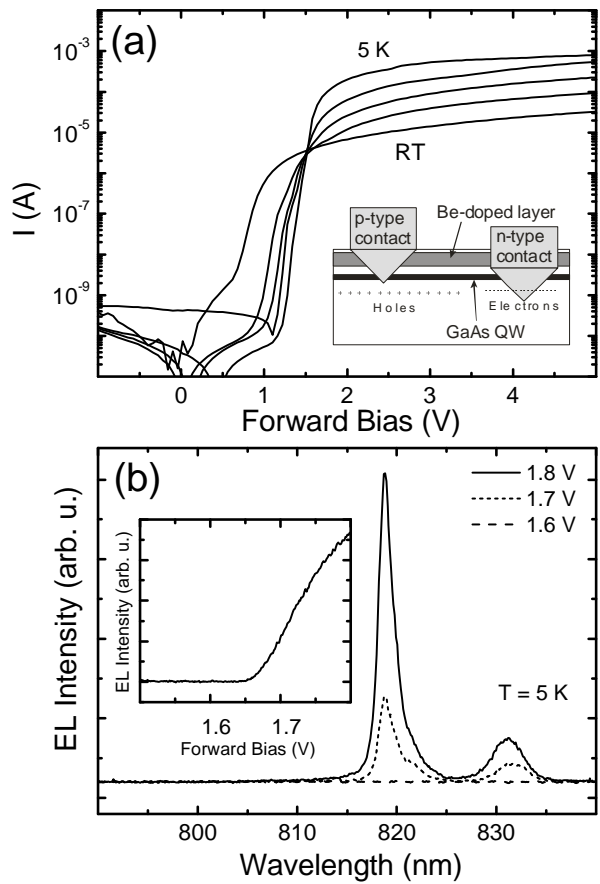}
\caption{(a) Current-voltage characteristics from room temperature
down to 5\,K. Inset: fabrication scheme of the n-p lateral junction. 
(b) Electro-luminescence spectra at different forward biases around 
the emission detecting threshold. Inset: Light-voltage characteristic 
of the planar LED at T = 5\,K.}
\label{fig1}
\end{figure}
The n-contact was shaped as a thin stripe placed perpendicular to the
SAW propagation direction, 250\,$\mu$m away from the p-contact.
 
SAWs propagating along the (0\={1}\={1})
crystal direction were generated by means of an interdigital
transducer (IDT) composed of 100 pairs of 200\,$\mu$m-long Al
fingers with 3\,$\mu$m periodicity ($\sim$1\,GHz resonance
frequency on GaAs). Transducers were fabricated at a distance of
800\,$\mu$m from the mesa by electron-beam lithography. 
The width of the n-type contact stripe (2\,$\mu$m) was
chosen of the same order of magnitude as the SAW wavelength
(3\,$\mu$m) in order to limit SAW damping due to the massive
metalization and SAW diffusion originating from non-uniform
penetration of the Ohmic contact and inhomogeneities in the etched
region. 

Several SAWLEDs were fabricated and studied, and all of them 
showed qualitatively similar characteristics. All the data shown
in the following refer to one representative device, for which the 
transport and optical properties of the junction, as well as the 
efficiency of the transducer, will be systematically analyzed.

The SAWLED was first electrically tested at low temperature to 
verify the proper formation of the lateral diode. Current-voltage 
curves (IVs) at different temperatures (from room temperature down 
to 5\,K) showed rectifying behavior and intersected at $\sim$1.5\,V 
(see Fig.\,1(a)). This value corresponds to the diode conduction 
threshold and it is consistent with the value expected for GaAs n-p 
junctions.
Emission properties were characterized by EL measurements at
low temperature (5\,K). Spectra as a function of bias were collected by
a cooled CCD after spectral filtering by a single-grating
monochromator. The EL spectra reported in
Fig.\,1(b) show a main peak at 818.7\,nm
(full width half maximum (FWHM) of 1.8\,nm) originating from radiative recombination within the
QW and a secondary peak at 831.2\,nm (FWHM 3.6\,nm) which originates from carbon impurities included in the
heterostructure material during growth\,\footnote{B. Hamilton, in
\emph{Properties of Gallium Arsenide}, edited by M. R. Brozel and
G. E. Stillman (INSPEC, London, England, 1996).}. The origin of 
such peaks was also verified by photo-luminescence 
measurements\footnote{PL spectra were obtained by
excitation of a region of the mesa with a red-light laser source
(653\,nm).}. Light intensity, calculated by
integration over the main EL peak from 810\,nm to 826\,nm, as a 
function of the forward bias applied to the junction reflects 
the rectifying behavior of the IV curves (inset of Fig.\,1(b)), 
showing a detection threshold of $\sim$1.65\,V.

The SAW resonance frequency was determined by measuring the power
reflected by the transducer as a function of excitation frequency.
The transducer frequency response displayed at 5\;K a pronounced dip at 987.5\;MHz, which is
consistent with the periodicity of the transducer, and had a FWHM of 2.4\;MHz.

The travelling electric field associated with SAWs can drive electrons into
the 2DHG when the junction is biased just below threshold \cite{cecchini-apl04, hosey-apl04}.
Indeed, we observed the light versus voltage curve to shift 
to lower biases by $\sim$10\,meV in presence of SAWs (Fig.\,2(a)).
\begin{figure}[!ht]
\includegraphics[width=8cm]{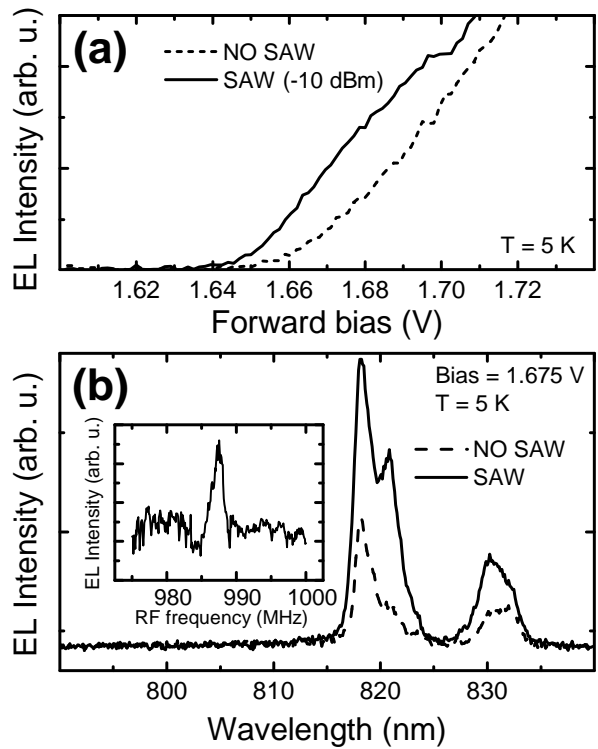}
\caption{(a) Light-voltage 
characteristic of the planar LED without SAW (dashed line) and with
SAW (solid line). SAW power and frequency were -10\,dBm and 987.5\,MHz
respectively. (b) Electro-luminescence 
spectra without SAW (dashed line) and in presence of a SAW
(-10 dBm, 987.5 MHz) (solid line). Inset: electro-luminescence 
as a function of RF-frequency (-9 dBm) at T = 5\,K. The junction 
was forward biased with 1.8 V.}
\label{fig2}
\end{figure}
The maximum shift was observed at -10 dBm power level. At 
higher levels the increased electron extraction is counterbalanced by
the spatial separation of electrons 
and holes trapped respectively in SAW minima and maxima, leading
to a progressive suppression of the EL signal.\cite{cecchini-apl04}

The emission spectra were changed by the SAW only in intensity and no spectral shift of the main peaks was observed (Fig.\,2(b)). 
Light-intensity data as a function of IDT excitation frequency 
demonstrate that emission intensification occurs
only within the transducer pass-band, as shown in the 
inset of Fig.\,2(b).

We also studied the SAW-induced EL signal by time-resolved measurements.
To this end the EL signal was spectrally filtered by a 
triple grating monochromator and detected by a single-photon APD 
module (Perkin Elmer SPCM-AQR-16). In order to obtain time-resolved EL
traces we used time-correlated photon-counting techniques.
The signal from the SPCM was directed toward the START input of a 
Becker \& Hickl SPC-600 computer board, while 
the STOP input was driven by a signal synchronous to that
used to generate the SAW.
A typical data trace is shown in Fig.\,3(a).
\begin{figure}[!ht]
\includegraphics[width=8cm]{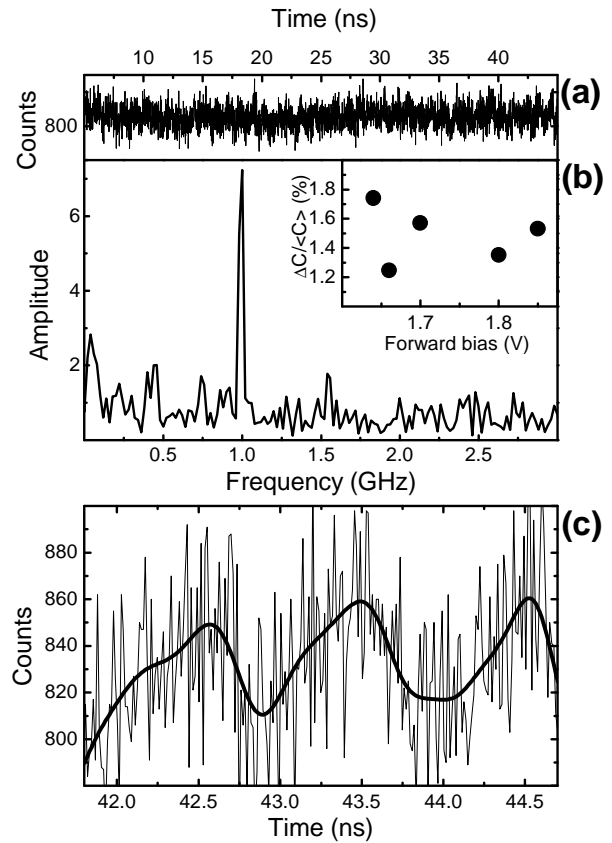}
\caption{(a) EL time evolution in presence of SAW 
(987.5\,MHz, -10\,dBm), at T = 5\,K and for a forward bias of 
1.85\,V. (b) Fourier transform of the top panel EL 
signal. Inset: Normalized oscillation intensity as a function of the forward bias. 
(c) Zoom of the time evolution of SAW-induced EL showed 
in (a) superimposed onto an FFT smoothing.}
\label{fig3}
\end{figure}
EL oscillations with period of about 1\,ns are present in the 
curve, reflecting the presence of the SAW modulating the junction 
potential. These oscillations are more clearly visible in 
Fig.\,3(c) where a magnified view of Fig.\,3(a) is displayed with a 
smoothed data trace.
This effect is further highlighted in the frequency 
domain of the signal. Fig.\,3(b) reports the 
Fourier transform of the EL signal. A single peak at the SAW frequency
clearly dominates the Fourier transform. 

Time-correlated measurements were made at different biases applied 
to the junction. The oscillation amplitudes were 
calculated from the Fourier transforms and normalized by the 
corresponding mean EL intensities. In the inset of Fig.\,3(b) we report the 
normalized oscillation intensity as a function of the forward bias. The oscillation amplitude 
is around 1.5\,\% of the signal and remains almost unchanged at different 
bias voltages, as expected for an exponential increase of the EL close 
to the conduction threshold.
Since the recombination time in our system is much less than 1 ns, 
determined by means of photoexcited-carrier lifetime measurements, 
the reason for the observed small oscillation amplitude originates
from other limiting factors.
Ideally the SAW wavefronts would get to the thin n-type contact without any 
distortion and perfectly parallel to the contact itself. But real 
devices introduce deviations from the picture described 
above. The presence of the mesa edges and of defects along the 
SAW path introduced during the mesa etching process constitute 
scattering centers which lead to modification of the acoustic 
wave-fronts. In addiction, annealed Ohmic contacts
have a strong corrugation. During the annealing process 
metal diffusion makes the junction edges irregular on the scale 
of the SAW wavelength. These two factors lead to electrons being 
extracted and recombining at different times depending on their 
position along the junction. The resulting oscillations are therefore 
averaged out, while the EL mean-intensity increase still remains present.
We believe that further optimization of the device geometry will lead to high 
contrast oscillations and is currently being investigated.

In conclusion, we fabricated n-p lateral junctions with interdigital 
transducers and studied the effect of SAWs on 
the device time-resolved emission properties. 
SAW-induced emission was first characterized by spectral measurements 
and light-voltage measurements. Electroluminescence was observed 
to increase in intensity in the presence of SAWs when 
the diode was biased near the conduction threshold. 
Time-resolved measurements of the electroluminescence in the presence 
of SAWs were carried out at several bias voltages, demonstrating 
modulation of the light intensity at the frequency of the 
SAW ($\sim 1$ GHz). The amplitude of the oscillation was found to 
be $\sim 1.5 \%$ of the total light intensity, almost 
independent from the bias applied to the junction. 

This work was supported in part by the European Commission through
the FET Project SAWPHOTON and by MIUR within FISR ``Nanodispositivi
ottici a pochi fotoni''.


\bibliography{050104_tSAWLED_cond-mat}

\end{document}